\documentclass[twocolumn,showpacs,superscriptaddress,preprintnumbers,amsmath,amssymb]{revtex4}

\usepackage{color}
\usepackage{graphicx}
\usepackage{dcolumn}
\usepackage{bm}

\begin{document}


\title{Orbital disorder induced by charge fluctuations in vanadium spinels}   

\author{Yasuyuki Kato}
\affiliation{Theoretical Division, Center for Nonlinear Studies, Los Alamos National Laboratory, Los Alamos, NM 87545, USA}

\author{Gia-Wei Chern}
\affiliation{Theoretical Division, Center for Nonlinear Studies, Los Alamos National Laboratory, Los Alamos, NM 87545, USA}
\affiliation{Department of Physics, University of Wisconsin, Madison, Wisconsin  53706, USA}

\author{K. A. Al-Hassanieh}
\affiliation{Theoretical Division, Center for Nonlinear Studies, Los Alamos National Laboratory, Los Alamos, NM 87545, USA}
\affiliation{Center for Nanophase Materials Sciences, Oak Ridge National Laboratory, Oak Ridge, Tennessee 37831, USA}

\author{Natalia B. Perkins}
\affiliation{Department of Physics, University of Wisconsin, Madison, Wisconsin  53706, USA}

\author{C. D. Batista}
\affiliation{Theoretical Division, Center for Nonlinear Studies, Los Alamos National Laboratory, Los Alamos, NM 87545, USA}

\date{\today}

\begin{abstract}
Motivated by recent experiments on vanadium spinels, $A$V$_2$O$_4$, that show an increasing 
degree of electronic delocalization for smaller cation sizes, we study the evolution of orbital ordering 
(OO) between the strong and intermediate-coupling regimes of a multi-orbital Hubbard Hamiltonian.
The underlying magnetic ordering of the Mott insulating state leads to a rapid suppression of
OO due to enhanced charge fluctuations along ferromagnetic bonds. 
Orbital double-occupancy is rather low at the transition point indicating that the system
is in the crossover region between strong and intermediate-coupling regimes
when the orbital degrees of freedom become disordered.
\end{abstract}


\maketitle

The steady interest in frustrated magnets with degenerate orbitals
 is driven by the continuous discovery of unusual magnetic and orbital orderings resulting from
an intricate interplay between frustration, lattice distortions and electron correlations.
A case in point is the family of vanadium spinels, $A$V$_2$O$_4$ ($A$ = Cd, Mg or Zn), whose
magnetic V$^{3+}$ ions reside on a pyrochlore lattice and contain
two electrons in their $t_{2g}$ $3d$-orbitals \cite{lee,reehuis,bella,onoda03,giovannetti}.
What makes this family particularly attractive is the  possibility of tuning the ratio between 
the electronic hopping, $t$, and the intra-orbital Coulomb  repulsion, $U$, by changing the cation size at $A$ sublattice~\cite{blanco}.

By using a strong-coupling approach,  Tsunetsugu and Motome 
found an antiferro-orbital (AFO) order consisting of alternating $d_{zx}$ and $d_{yz}$ orbitals along both $[1,0,\pm1]$ ($zx$) and$[0,1,\pm1]$ ($yz$) directions \cite{tsunetsugu}. However,  
AFO is incompatible with the crystal symmetry $I4_1/amd$ extracted from
neutron scattering (NS) and x-ray diffraction experiments \cite{lee,reehuis,bella}.
Tchernyshyov \cite{tcher} proposed that AFO is suppressed
by a strong spin-orbit (SO) interaction \cite{tcher,dimatteo,maitra}.
Although there is no reliable data on the SO coupling for V$^{3+}$ ions, free ion measurements \cite{abragam}
and {\em ab initio} calculations \cite{mizokawa} indicate that it may be comparable to the exchange energy.
However, recent NS measurements on MgV$_2$O$_4$ \cite{bella} detected a  small spin gap and highly dispersive magnetic excitations that are at odds with strong SO coupling \cite{perkins}. 
More recent experimental studies of the $A$V$_2$O$_4$ family show that none of these compounds satisfy the phenomenological  Bloch's 
equation \cite{Bloch66}
$  \partial \ln{T_N}/ \partial \ln{V} \simeq 3.3$,
that must hold in the strong-coupling limit $t/U \ll 1$ \cite{blanco} ($T_N$ and $V$ are  N{\'e}el temperature and volume). Moreover,  the  N{\'e}el temperature of ZnV$_2$O$_4$ decreases with pressure  and  transport measurements reveal that MgV$_2$O$_4$ and ZnV$_2$O$_4$ have  small charge gaps \cite{pardo}.
These measurements clearly indicate that a comprehensive  study of the spinel Vanades requires an approach that can interpolate 
between the strong and intermediate-coupling regimes.

\begin{figure}
\vspace*{-0.8cm}
\vspace{0.5cm}
\includegraphics[angle=0,width=9cm]{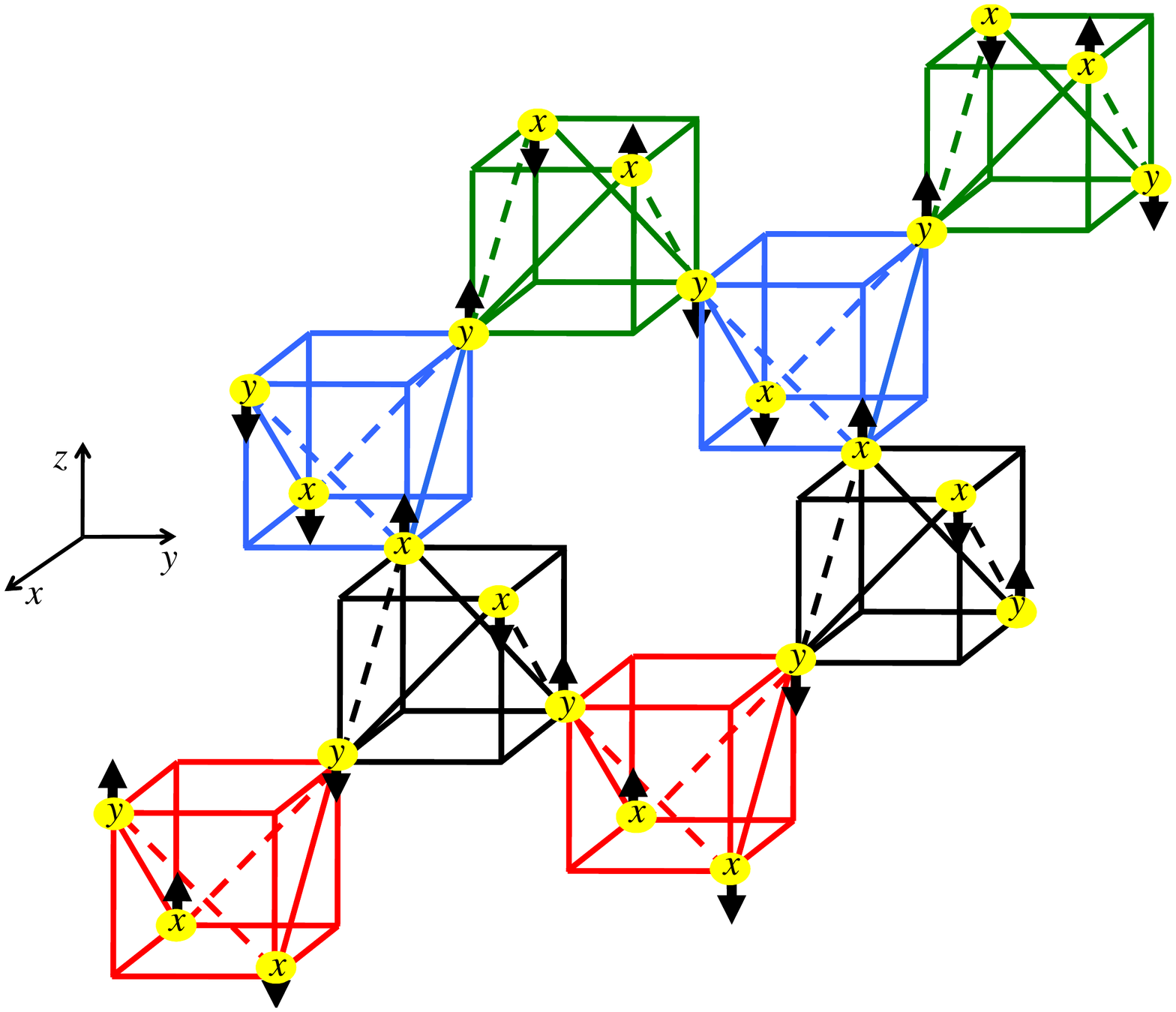}
\vspace*{-0.9cm}
\caption{(Color online) Pyrochlore lattice of V$^{3+}$ ions in AV$_{2}$O$_4$. The solid diagonal lines are FM bonds along
$zx$ and $yz$ directions. These ``strong'' bonds form zig-zag chains described by ${\bar H}$.
The dashed lines are the ``weak'' AFM bonds that introduce inter-chain orbital coupling. The arrows indicate the spin ordering
favoured by a combination of the intra-chain FM coupling
and inter-chain AFM coupling induced by 
bonds oriented along the $xy$ direction.
The letters $x$ (for $zx$) and $y$ (for $yz$) indicate the OO
that is stable deep inside the Mott regime \cite{tsunetsugu}.}
\label{3dlat}
\end{figure}

In this Letter we demonstrate that the larger charge fluctuations of the intermediate-coupling regime
play a crucial role for suppressing OO in MgV$_2$O$_4$ and ZnV$_2$O$_4$.  The observed
magnetic ordering breaks the equivalency between bonds and the strong Hund's coupling
results in a lower energy barrier for ferromagnetic (FM) bonds. Since the FM bonds  form  zig-zag chains spiraling along the $z$ direction (see Fig.~\ref{3dlat}), 
charge fluctuations become stronger along these chains. 
We argue that it is essential to keep double occupied states in the low-energy effective theory  to account for the
lower energy barrier of FM bonds. In fact, we show that double occupied states of isolated zig-zag chains
are domain walls of a 1D quantum Ising model. These domain walls are confined in the
orbitally ordered phase.  As $t/U$ increases, the zig-zag chain undergoes a quantum phase transition to a para-orbital (PO) \cite{note1} state via proliferation of  domain walls. We find that this transition takes place at the crossover between the intermediate and strong-coupling regimes. 
This phenomenon cannot be captured by a strong-coupling approach because double-occupied states are projected out from the low-energy Hilbert space.

We first review experimental results on vanadium spinels.  A structural transition occurs at a temperature $T_s \approx 95$K 
for $A$ = Cd \cite{onoda03,giovannetti},  
$T_s \approx 51$K for $A$ = Zn \cite{reehuis}, and   $T_s \approx  65$K for $A$ = Mg \cite{bella}, which lowers
the crystal symmetry from cubic $Fd\bar 3m$ to tetragonal $I4_1/amd$
and leads to uniform flattening of VO$_6$ octahedra with
$c<a=b$ . This distortion leads to a partial ferro-orbital (FO) ordering in which
the lower-energy $d_{xy}$ orbital is occupied at every site, while the second electron can occupy either the $d_{zx}$
or $d_{yz}$ orbitals.  Antiferromagnetic correlations
develop below $T_s$ along chains
parallel to $[1,\pm1,0]$ ($xy$)  directions.
However, 3D magnetic ordering  only sets in below  a lower  Neel temperature 
due to frustration in the inter-chain coupling. The ordering  wave-vector is $\mathbf q = 2\pi(0,0,1)$ and the corresponding 
spin pattern is $\uparrow\uparrow\downarrow\downarrow$ along 
chains parallel to the $yz$ or $zx$ directions.  
This ordering leads to zig-zag FM chains spiraling about the $z$-axis  (Fig.~\ref{3dlat}).

{\it The Model.} We start by considering a low-energy $t_{2g}$ Hamiltonian
$H = H_{cf} + H_U + H_t + H_{so}$. The first term,
$H_{cf} = -\Delta \sum_{j} n_{j\, xy}$,
describes the crystal field 
splitting due to the Jahn-Teller distortion at $T < T_s$,
where $n_{j\, xy}$  is the electron number for the $d_{xy}$ orbital of site  $j$.
We also assume a value of $\Delta > 0$ that is large enough to localize one electron in the  $d_{xy}$ 
orbital.
$H_U$ contains the terms originated from the Coulomb repulsion between electrons in the same ion.
When restricted to the 
$n_{j\, xy} = 1$ subspace, $H_U$ reads
\begin{eqnarray}
\label{eq:H0}
& & H_U = \sum_{j,\mu} \left[ - 2 J {\bf S}_{j\mu} \cdot {\bf S}_{j\,xy}
+ U n_{j \mu \uparrow} n_{j \mu \downarrow} \right] \nonumber \\
& &\,+ (U-2J) \sum_{j, \mu \neq \nu} n_{j \mu \uparrow} n_{j \nu \downarrow}
+ \frac{U -3J}{2} \sum_{j, \alpha, \mu \neq \nu} n_{j \mu \alpha} n_{j \nu \alpha} 
\nonumber \\
& &\,+ J \sum_{j, \mu \neq \nu} \Bigl[d^{\dagger}_{j\mu\uparrow}
d^{\dagger}_{j\mu\downarrow} d^{\;}_{j\nu\downarrow} d^{\;}_{j\nu\uparrow}
- d^{\dagger}_{j\mu\uparrow} d^{\;}_{j\mu\downarrow} d^{\dagger}_{j\nu\downarrow}
d^{\;}_{j\nu\uparrow}\Bigr]. \nonumber \\
\end{eqnarray}
Here $U$ denotes the Coulomb repulsion between electrons occupying the same 
orbital and $J$ is the Hund's coupling constant \cite{dagotto01}. 
$\mu, \nu=\{zx, yz\}$ are orbital indices, while $\alpha, \beta = \uparrow$, $\downarrow$ are spin indices.
Finally, $n_{j \mu \alpha}= d^{\dagger}_{j\mu\alpha} d_{j\mu\alpha}$, $n_{j \mu} = \sum_{\alpha} n_{j \mu \alpha}$, and
$ {\bf S}_{j\mu}=\frac{1}{2} \sum_{\alpha,\beta} 
d^{\dagger}_{j\mu\alpha} \bm\sigma_{\alpha\beta}d^{\;}_{j\mu\beta}$,
where $\bm\sigma = (\sigma^x, \sigma^y, \sigma^z)$ is a vector of Pauli matricies.
The kinetic energy terms are
\begin{equation}
\label{eq:Ht}
H_t =  \sum_{jj'} \sum_{\mu, \nu, \alpha} t^{\mu \nu}_{jj'}
(d^{\dagger}_{j\mu\alpha}\, d^{\phantom{\dagger}}_{j'\nu \alpha} + {\rm H. c.})
\end{equation}
We assume that the transfer matrix is diagonal in
the $t_{2g}$ manifold and that the hopping integral is dominated by the $dd\sigma$ contribution: 
$t^{\mu \nu}_{jj'} = t^{\mu\mu}_{jj'} \delta_{\mu,\nu}$.

Finally, the effective SO contribution $H_{so}$ is obtained by projecting the original SO interaction,
$\lambda {\bf L} \cdot {\bf S}$, onto the doublet  of $\{d_{zx}, d_{yz}\}$ orbitals \cite{horsch03}:
\begin{eqnarray}
\label{eq:Hso}
H_{so } = i\lambda \sum_{j \alpha} \sigma^z_{\alpha\alpha} 
\left(d^{\dagger}_{j\, zx\,\alpha} d^{\phantom{\dagger}}_{j\, yz\,\alpha} 
- d^{\dagger}_{j\, yz\,\alpha} d^{\phantom{\dagger}}_{j\, zx\,\alpha}\right).
\end{eqnarray}
The SO coupling also contains terms, like $\lambda d^{\dagger}_{j\, xy \uparrow} d_{j\, \mu\, \downarrow}$,
which mix the $d_{xy}$ with 
$d_{zx}$ or $d_{yz}$ orbitals. Since these terms are of
order $\lambda/\Delta$, they will be neglected in the following discussion.

{\it A single helical chain ($\lambda=0$ limit).} 
We now consider a single helical chain that propagates along $z$-direction with alternating
$zx$ and $yz$ bonds.
(We use the short notation ``$\mu$-bond''  for bonds oriented along the $\mu$ direction, where $\mu=\{ xy, yz, zx\}$.)
The hopping matrix elements along each helical chain are 
$t^{zx,zx}_{j,j+1}=t$ and $t^{yz,yz}_{j,j+1}=0$ for $zx$-bonds,
$t^{yz,yz}_{j,j+1}=t$ and $t^{zx,zx}_{j,j+1}=0$ for $yz$-bonds,
while there is no hopping between $d_{xy}$ orbitals. 
The resulting single-chain hopping Hamiltonian is
\begin{eqnarray}
t \sum_{j \in {\rm odd},\, \alpha} \!\! \left(d^{\dagger}_{j+1\, zx\, \alpha} d^{\phantom{\dagger}}_{j\, zx\, \alpha}
+ d^{\dagger}_{j-1\, yz \alpha} d^{\phantom{\dagger}}_{j\, yz\, \alpha} + {\rm H.c.}\right).
\end{eqnarray}
The total charge in each pair of orbitals connected by a finite hopping amplitude  
is conserved for $\lambda=0$. This local U(1) invariance of $H(\lambda=0)$
makes the model quasi-exactly solvable. 
For realistic Hamiltonian parameters, the ground state of $H(\lambda=0)$ is always in the fully
polarized subspace ${\cal S}$  with exactly one electron per bond.
The projection of  ${H}(\lambda=0)$ onto this invariant subspace is mapped into a quantum Ising model (QIM):
\begin{eqnarray}
\label{1DIsing}
P_{{\cal S}} {H}(\lambda=0) P_{{\cal S}} = 
- {\cal J} \sum_j \left[ \tau^z_{j,j+1} \tau^z_{j+1,j+2} - g \tau^x_{j,j+1} \right], 
\end{eqnarray}
up to a constant $C= N_s (U - J)/4$. Here  ${\cal J}=(U-3J)/4$, $g = 4 t /(U -3J)$, and $N_s$ is the total number of V$^{3+}$ ions in the chain. 
The Ising variable $\tau_{j,j+1}^z$ is
equal to $1$ if an electron occupies the right site ($j+1$) of the bond and $-1$ if it occupies the left site ($j$).

\begin{figure}[!htb]
\includegraphics[angle=0,width=8cm]{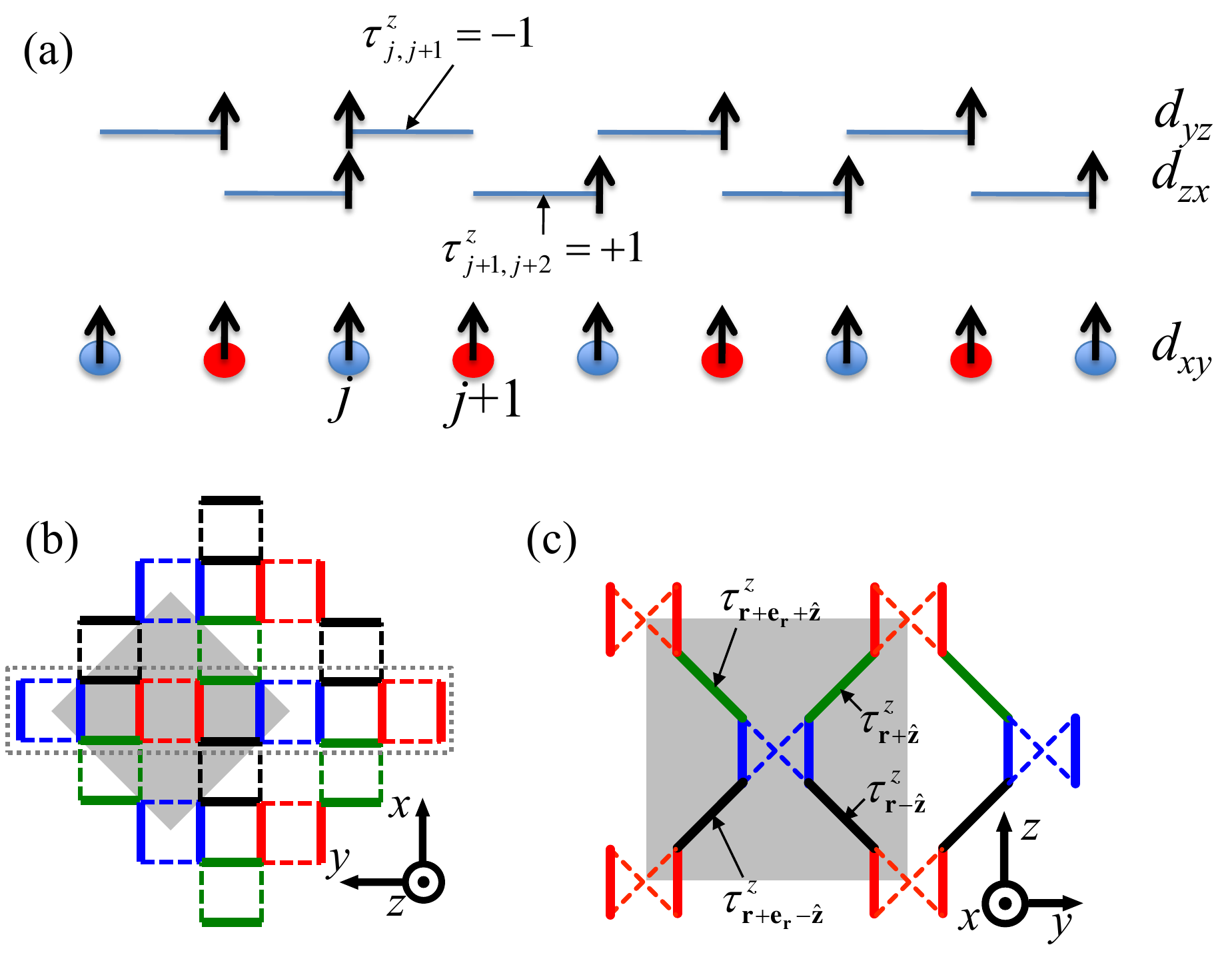}
\vspace*{0.0cm}
\caption{(Color online) (a) Mapping between ${H}_{\lambda=0}$ and the QIM. The even and odd sublattices are indicated
with blue and red circles respectively. (b) and (c) show the projections of the pyrochlore lattice and the helicoid Ising chains (solid lines) on
the $xy$ and $yz$ planes, respectively. }
\label{Ising}
\end{figure}

The operator that is associated with the local orbital order parameter, 
$n_{j\,zx}-n_{j\,yz}$, has the following
expression in terms of the Ising variables:
\begin{eqnarray}
n_{j\,zx}-n_{j\,yz} &=& \pm \frac{\tau^z_{j \pm 1,j\pm1+1}+\tau^z_{j,j+1}}{2},
\label{orb}
\end{eqnarray}
where the $+$ ($-$) sign holds for odd (even) values of $j$. The 1D QIM is
exactly solvable and the ground state has  FM ordering  for 
$U-3J \geq 4 t$.
The corresponding order parameter is:
$
\langle \tau^z_{k=0} \rangle = \sum_j \langle \tau^z_{j,j+1} \rangle.
$
According to Eq.~\eqref{orb}, FM ordering of  the Ising variables corresponds to 
AFO ordering of the original variables:
\begin{equation}\label{orderparameter}
{\cal O}_{\pi} = \frac{1}{L} \sum_j e^{i \pi j} \langle n_{j\,zx} - n_{j\,yz} \rangle  = -\langle \tau^{z}_{k=0} \rangle,
\end{equation}
where $L$ is the number of sites in the helical chain.
Therefore, the quantum phase transition between the FM and paramagnetic (PM) states of the Ising variables corresponds
to AFO-PO transition in terms of the original variables.

The quantum critical point (QCP) occurs at 
$t=t_c=(U-3J)/4$ or $|g|=1$.
The exact value of the nearest-neighbor correlator at the QCP is
$
\langle \tau^z_{j,j+1}\tau^z_{j+1,j+2} \rangle_c = 2/{\pi},
$
which implies a rather low probability of double-occupancy:
$
\langle n_{j\,zx} n_{j\,yz}\rangle_c = \frac{1}{4} \left(1-\langle \tau^z_{j,j+1} \tau^z_{j+1,j+2} \rangle_c \right)\simeq 0.09.
$
This means that the transition to the PO state occurs far from the covalent regime and
the inter-chain orbital coupling can be treated as a perturbation.

{\it Coupled Chains.}
Here we will assume that the effective AFM coupling between chains stabilizes the magnetically ordered state shown in Fig.\ref{3dlat}. This assumption is supported 
by unbiased numerical simulations of the three-band Hubbard model that will be presented elsewhere \cite{kato12}. In addition, this magnetic ordering is stable near the itinerant \cite{Chern11} and strong-coupling limits \cite{tsunetsugu} indicating that it remains stable over the whole Mott phase. 
Since charge fluctuations are weaker across AFM bonds (the barrier is $U$ instead of $U-3J$), the coupling
between neighboring helical Ising chains [Fig. 2(c)] will be approximated by using a Kugel-Khomskii 
Hamiltonian \cite{kk,tsunetsugu,dimatteo}. There are two contributions. The first contribution comes from  exchange 
between electrons localized in the $d_{xy}$ orbitals and leads to a pure AFM spin coupling:
\begin{eqnarray}
	H_{\rm spin} = J_{S} \sum_{(ij)} \mathbf S_i \cdot \mathbf S_j,
\end{eqnarray}
where $J_{S} =  \frac{t^2}{U}\frac{1+\eta}{1+2\eta}$ is the spin exchange constant, 
$\eta = J/U$, $(ij)$ denotes an $\pm xy$-bond which connects two sites belonging to nearest-neighbor
Ising chains, and ${\bf S}_j=\sum_{\gamma}{\bf S}_{j\gamma}$. The above AFM coupling between helical chains
is unfrustrated and leads to the $\mathbf q=2\pi(001)$ 3D magnetic order depicted in Fig.~\ref{3dlat}.

The second contribution comes from orbital exchange through the antiferromagnetic $zx$ and $yz$ bonds (dashed bonds in Fig.~\ref{3dlat})
connecting nearest-neighbor Ising chains. The  small  probability of double occupancy induced by inter-chain hopping processes justifies our
perturbative treatment of these terms. The resulting inter-chain orbital Hamiltonian is 
\begin{eqnarray}\label{interorbital}
H_{\rm orb} =  \sum_{\langle ij \rangle } K_1 n_{i \mu} n_{j \mu} +  K_2  \left[ n_{i \mu} (1-n_{j \mu}) + (1-n_{i \mu}) n_{j \mu} \right],  \,\,
\end{eqnarray}
where 
$\mu = zx$ ($yz$) when $\langle ij \rangle$ is a $zx$ ($yz$)-bond,
$K_1 = -2\frac{t^2}{U}\frac{1+\eta}{1+2\eta}$ and 
$K_2 = -\frac{t^2}{U}\frac{1-2\eta}{1-3\eta}$ denote
the FO and AFO couplings.

\begin{figure}[!htb]
\vspace*{-0.0cm}
\hspace*{-0.0cm}
\includegraphics[angle=0,width=8.0cm]{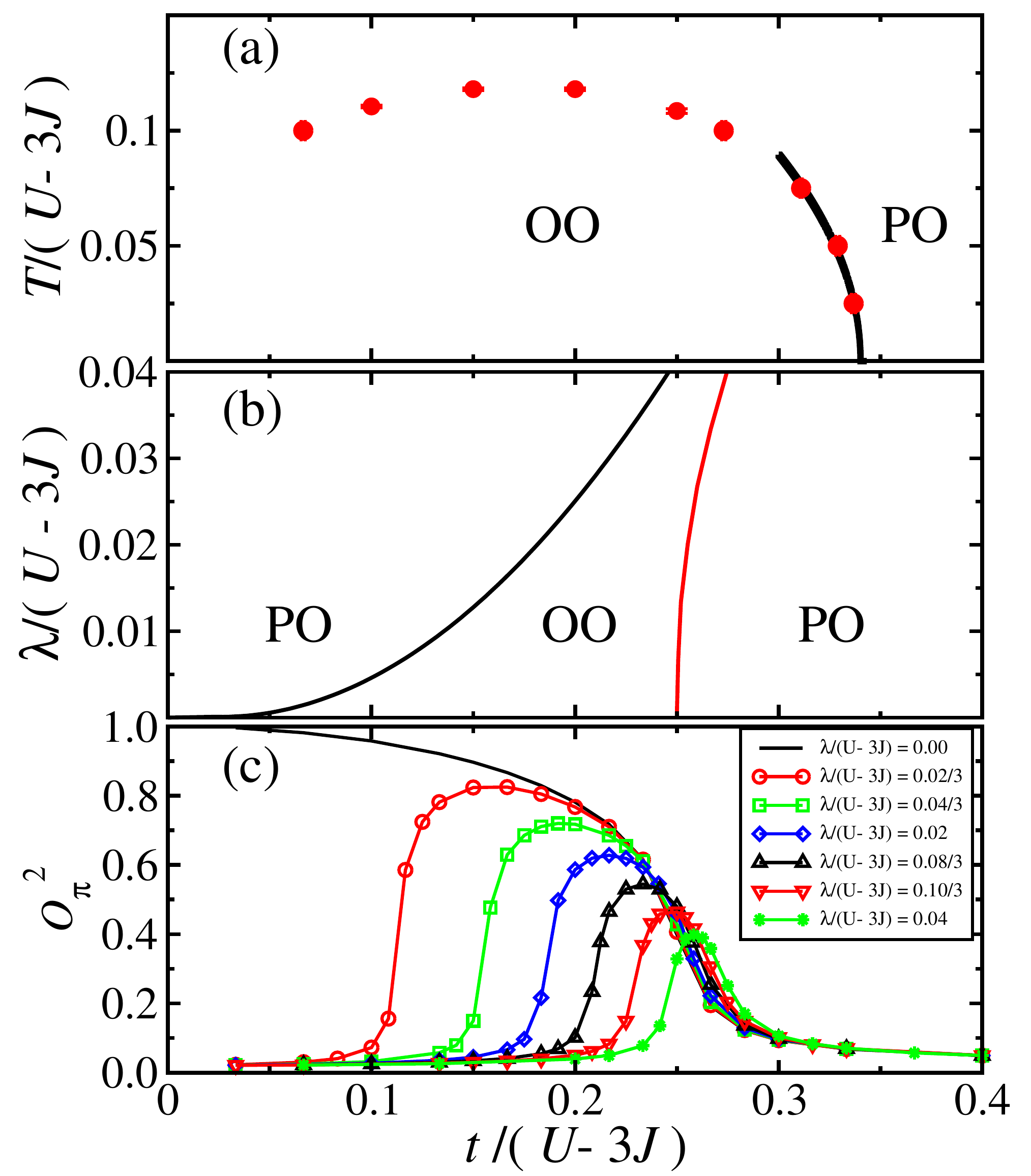}
\caption{(Color online) (a) Phase diagram of the three-dimensional quantum Ising model Eq.~(\ref{eq:Heff}) for $\eta=0.229$.
(b) $T=0$ ($\lambda, t$) phase diagram of the single helical chain Hamiltonian $H$ obtained with DMRG applied to
chains of 48 unit cells. (c) Square of the staggered orbital order parameter as a function of $t$.}
\label{interchain}
\end{figure}

A 3D effective Ising Hamiltonian can be easily obtained from Eqs.\eqref{1DIsing} and \eqref{interorbital}. The intra-chain term  is given by
Eq.(\ref {1DIsing}), while the inter-chain coupling  is obtained by expressing  the orbital occupation operators of Eq. (\ref{interorbital})
in terms of the Ising variables
$n_{j \mu} = (1 \pm \tau^{z}_{j,j+1})/2$.
However, we should recall that the Ising operators are bond variables defined on a dual lattice (see Fig. 2(a)). Therefore,
we introduce the bond coordinates $\mathbf r = (m, n, j)$ 
to define the full dual lattice, including the $zx$ and $yz$-bonds connecting different helical chains.
The last coordinate $j$ denotes the position of the bond on its  helical chain, while
$m$ and $n$ correspond to the $(x, y)$ chain coordinates. (See Fig.~\ref{Ising}.)
The resulting quantum Ising Hamiltonian is
\begin{eqnarray}
\label{eq:Heff}
H_{\rm eff} &=&  -\frac{U-3J}{4}\sum_{\bf r} \left(\tau^z_{{\bf r} + {\hat{\bf z}}}\,\tau^z_{\bf r}
	- g\,\tau^x_{\bf r} \right) \\
 & & -\eta\frac{ t^2}{U} \sum_{\mathbf r}  \left( \tau^z_{{\bf r} + {\bf e}_{\bf r} - {\hat{\bf z}}}  \,
\tau^z_{{\bf r} + {\hat{\bf z}}} + \tau^z_{{\bf r} + {\bf e}_{\bf r} + {\hat{\bf z}}} \,
\tau^z_{{\bf r} - {\hat{\bf z}}} \right), \nonumber
\end{eqnarray}
where 
$\mathbf e_{\mathbf r} =(\pm 1, 0, 0)$ and $(0, \pm 1, 0)$  are
vectors connecting the site $\mathbf r$ to its neighbors. The inter-chain orbital coupling is much weaker than the intra-chain coupling and 
both are FM.  In the  large $U/t$ limit ( deep inside the MI phase), the FM coupling between Ising variables  leads to the  AFO order along dashed $zx$ or $yz$ bonds connecting different helical chains (see Fig.~\ref{3dlat}).
This AFO alignment  is in contradiction with naive expectations  based on the a single-bond analysis.
$|K_1| > |K_2|$ for finite $\eta$  and the orbital superexchange favors a FO configuration the dashed $zx$- and $yz$-bonds. However, the state with FO alignment along dashed  bonds is frustrated 
because half of those bonds would contain pairs of occupied orbitals that are not connected by a finite hopping amplitude.
In contrast,  the energy gain is the same for every bond  of the  AFO order shown in Fig.~\ref{3dlat}.   

Fig.~\ref{interchain}(a) shows the thermodynamic phase diagram of  $H_{\rm eff}$ (\ref{eq:Heff}) obtained from  quantum Monte Carlo (QMC) simulations on lattices containing up to 8$\times$8$\times$40 unit cells (20480 sites). As in the 1D case, the transition between OO and PO occurs in the crossover region between the strong and intermediate-coupling regimes. 
As expected,  the ordering temperature, $T_{\rm OO}$,  increases with $U$.

{\it Finite SO coupling.}
To compare two different mechanisms for suppression of the AFO ordering,
we return to the original Hamiltonian $H$ on a  single helical chain and  quantify the effect of a finite SO interaction.
We apply the density matrix renormalization group (DMRG) method to a chain of
16 sites and verify that  the ground state of $H$ is still a fully polarized ferromagnet for $\lambda \leq 0.05(U-3J) $ in the entire regime parameters that we have been considering here. 
We  project $H$ into the fully polarized subspace ${\cal S}$ and split each site of the helical chain into two single-orbital sites. Then we arrange the orbitals in a one-dimensional array
$d_{j\,zx}, d_{j\,yz}, d_{j+1\,yz}, d_{j+1\,zx},...$  and identify each orbital with an effective site $l$.   
The result is an effective  spinless fermion model with alternating hopping and nearest-neighbor repulsion:
\begin{eqnarray}
P_{{\cal S}} {H} P_{{\cal S}} \!\!\! &=& \!\!\!
\sum_l t (c^{\dagger}_{2l-1} c^{\;}_{2l} + {\rm H. c. })+ (U -3J) n_{2l} n_{2l+1}
\nonumber \\
&+& i \lambda (c^{\dagger}_{2l} c^{\;}_{2l+1} - c^{\dagger}_{2l+1} c^{\;}_{2l} ).
\end{eqnarray}
The ground state  of $P_{{\cal S}} {H} P_{{\cal S}}$ is obtained by applying DMRG to a chain of 48 sites. 
The resulting ($\lambda, t$) quantum phase diagram   and the  AFO  order parameter 
${\cal O}_{\pi}$ [see Eq.~(\ref{orderparameter})] are presented in Fig.~\ref{interchain}(b) and (c), respectively. 
It is clear that  SO coupling and charge fluctuations effectively suppress the AFO order in different parts of the phase diagram.  
SO coupling $\lambda$ is very effective for suppressing the AFO deep inside the MI (large $U$) regime.
A  small SO coupling of about 6$\%$ of the Coulomb energy drives the system into the PO state because  
$\lambda$ competes against a super-exchange energy scale  of order $t^2/U$ that stabilizes the AFO order \cite{tcher}.
On the other hand,  SO coupling has  little effect  in the vicinity of the QCP because 
the competing energy scale that determines the strength of the charge  fluctuations is of order $t $. 
We should emphasize that although Fig.~\ref{interchain} includes a region deep inside the MI regime, our approach is  quantitatively correct only  near the QCP  that separates the OO and PO phases. 

In summary, our results offer a new perspective for understanding the electronic properties of the vanadium spinels $A$V$_2$O$_4$.
While CdV$_2$O$_4$ seems to be not too far from the localized or strong-coupling regime, it is not clear if the magnetic ordering is 
accompanied by OO. Different experimental probes indicate that MgV$_2$O$_4$ and ZnV$_2$O$_4$ are well inside the intermediate-coupling regime \cite{blanco,pardo,Kismarahardja11}.
MgV$_2$O$_4$ and ZnV$_2$O$_4$ exhibit the same type of $\uparrow \uparrow \downarrow \downarrow$ magnetic ordering and there is no evidence 
of OO down to the lowest accessible temperatures. 
According to our calculations, the SO interaction is very  effective for suppressing OO in the localized regime
relevant for  CdV$_2$O$_4$ \cite{tcher}. However, the lack of OO in the intermediate-coupling regime relevant for MgV$_2$O$_4$ and ZnV$_2$O$_4$ is many driven by
charge fluctuations and basically insensitive to the magnitude of the SO interaction.  While  SO still contributes to the rather large suppression of the V$^{3+}$ moment in the three compounds
(1.19$\mu_B$ in CdV$_2$O$_4$, 0.63 $\mu_B$ in ZnV$_2$O$_4$ and 0.47$\mu_B$ in MgV$_2$O$_4$),
we attribute the significantly smaller values observed in ZnV$_2$O$_4$ and MgV$_2$O$_4$ to the same charge fluctuations that suppress the OO.

ZnV$_2$O$_4$ and MgV$_2$O$_4$ have very similar lattice parameters \cite{blanco}. 
The estimated value of $t$ for the cubic phase of  ZnV$_2$O$_4$ with lattice
parameter 2.97~{\AA}  is $t \simeq 0.35$ eV \cite{takubo06}. According to our results, the OO should disappear completely for  $U-3J \gtrsim 1.2$ eV.
If we assume that $U \simeq 3.5$ eV \cite{pardo} and $J \simeq 0.8$ eV \cite{tsunetsugu,maitra}, 
OO should be  completely suppressed  in agreement with  experimental observations.
We note that the PO phase  found  for $t \geq 0.3 (U-3J)$ is similar
to the state obtained from an {\it ab initio} itinerant approach \cite{pardo}.
However, the bond order parameter associated with the $\uparrow \uparrow \downarrow \downarrow$ magnetic ordering is much weaker close to the QCP  than in the itinerant 
regime.  In other words, the  lattice distortion induced by the bond ordering   (FM bonds become shorter than the AFM ones) near the QCP should be much smaller  
than the value reported in Ref.~\cite{pardo}.  This could explain why recent NS measurements
have not observed the strong dimerization predicted in Ref.~\cite{pardo}. This may also be the reason why  {\it ab initio} calculations overestimate the electric polarization induced 
by the same bond  ordering in CdV$_2$O$_4$ \cite{giovannetti}. We believe that much better quantitative agreement  can be obtained from an intermediate-coupling treatment,
like the one presented here, that incorporates the coupling to the lattice degrees of freedom.

We are grateful to D. Khomskii, F. Rivadulla and V. Pardo for many useful discussions.
Work at the LANL was performed under the auspices of the
U.S.\ DOE contract No.~DE-AC52-06NA25396 through the LDRD program. A.E.F.
thanks NSF for support through Grant No. DMR-0955707. G.W.C. acknowledges the support
of ICAM and NSF grant DMR-0844115. N.P. acknowledges the support
of NSF grant DMR-1005932.

\end{document}